\documentclass[
reprint,
superscriptaddress,
amsmath,amssymb,
aps,
prl,citeautoscript,
]{revtex4-1}
\usepackage{color}
\usepackage{gensymb}
\usepackage{graphicx}% Include figure files
\usepackage{dcolumn}% Align table columns on decimal point
\usepackage{bm}% bold math
\usepackage{hyperref}% add hypertext capabilities
\usepackage{epstopdf}
\usepackage{float}

\begin{document}

\preprint{APS/123-QED}
\title{Rheology of Active Polymer-like \textit{T. Tubifex} Worms}

\author{A. Deblais}
\affiliation{Van der Waals-Zeeman Institute, Institute of Physics, University of Amsterdam, 1098XH Amsterdam, The Netherlands.}
\author{S. Woutersen}
\affiliation{Van 't Hoff Institute for Molecular Sciences, University of Amsterdam, Science Park 904, 1098XH Amsterdam, The Netherlands.}
\author{D. Bonn}
\affiliation{Van der Waals-Zeeman Institute, Institute of Physics, University of Amsterdam, 1098XH Amsterdam, The Netherlands.}

\date{\today}
            
\begin{abstract}
Of all complex fluids, it is probably the rheology of polymers we understand best. In-depth insight into the entanglement and reptation \cite{DeGennes1979,Doi1986} of individual polymers allows us to predict for instance the shear-thinning rheology and the behaviour in virtually any flow situation of practical importance \cite{Munk1989}. The situation is markedly different when we move from \emph{passive} \cite{LeDuc1999,Perazzo2017} to \emph{active} polymers where the coupling of filament activity, hydrodynamic interactions, and conformations open the way to a plethora of novel structural and dynamical features \cite{Liverpool2001,Marchetti2013,Winkler2017,Martin-Gomez2018,Martin-Gomez2019}. Here we experimentally study the rheology of long, slender and entangled living worms (\textit{Tubifex tubifex} \cite{Lazim1986}) and propose this system as a new type of active polymer. Its level of activity can be controlled by changing the temperature or by adding small amounts of alcohol to make the worms temporarily inactive. We find that (i) shear thinning is reduced by activity, (ii) the characteristic shear rate for the onset of shear-thinning is given by the time scale of the activity, and (iii) the low shear viscosity as a function of concentration shows a very different scaling from that of regular polymers. Our study paves the way towards a new research field of `living polymers'.
\end{abstract}

\pacs{Valid PACS appear here}% PACS
\maketitle

Active systems consist of interacting agents that are able to extract energy from the environment to produce sustained motion. The local conversion of energy into mechanical work drives the system far from equilibrium, yielding new dynamics and phases \cite{Zhang2010,Fodor2018}. Understanding the non-equilibrium statistical mechanics of such active systems is a major challenge, both experimentally and theoretically. While theoretically, much progress has been made recently, the number of experimental systems is still very limited \cite{Liverpool2001,Chen2007,Sokolov2009,Giomi2010,Koch2011,Zhou2013,Lopez2015,Tennenbaum2016} and often restricted to a relatively small number of simple entities such as driven colloidal particles \cite{Hatwalne2004, Linden2019, Delphine2019}. We study here the collective rheological behaviour of a system that is active and in structure greatly resembles polymer solutions and melts, a system eminently adapted to a statistical mechanical description that typically forms the basis of polymer models. Tubifex worms \cite{Lazim1986}, also called sludge worms, represent a simple and affordable system to study active polymer rheology. The worms are readily available in large quantities: they are commonly used as food for aquarium fish, and available in most pet shops. They have a typical length between 10 and 30~mm and a typical width of 0.2-0.4~mm (Supplementary Information section I.1).

In order to compare our living worms with active polymers, we first turn to a well-known phenomenon in polymer flows, which is that of shear thinning. It is due to fact that the thermal fluctuations of polymer chains tend to randomize their conformation, while the flow tends to orient and stretch them. The higher the flow rate, the more oriented the polymers, and the smaller the flow resistance. Shear thinning can also be due to entangled polymer chains getting disentangled as a result of the flow, again resulting in increased orientation and hence a smaller flow resistance with increasing shear rate.

\begin{figure*}
\includegraphics[width=16cm]{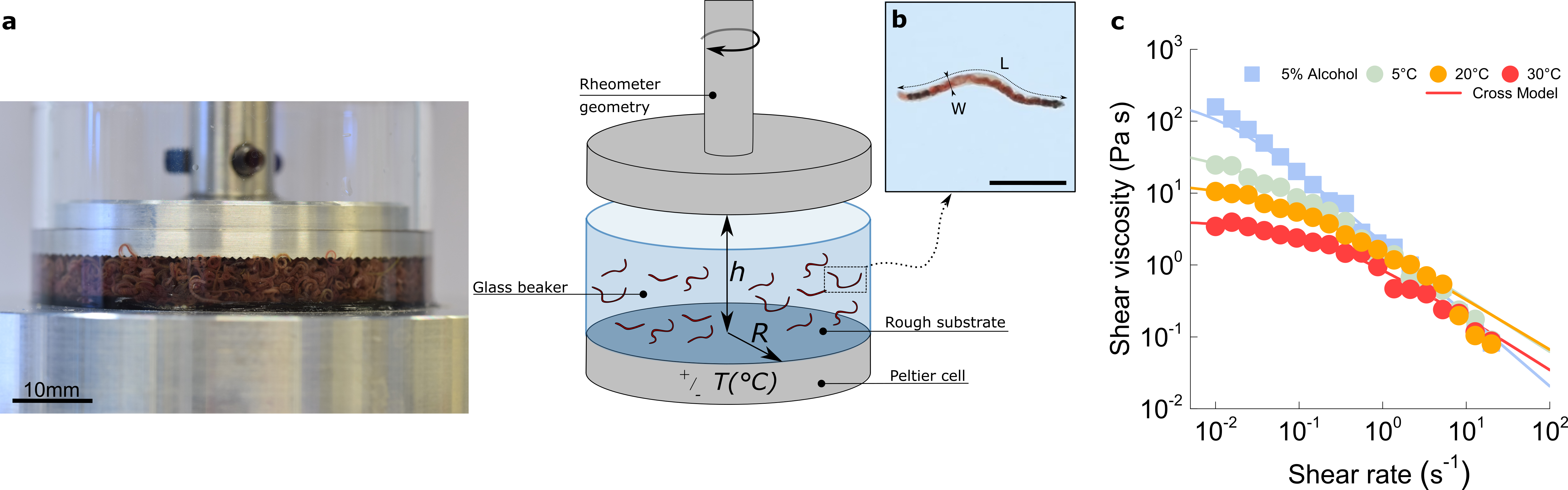}
\caption{\textbf{$\vert$ Rheology experiments of living polymers.} 
\textbf{a} Picture and schematics of the custom-designed rheology cell (see Supplementary Video 1) used for the measurements (not to scale). The cell is mounted on a Peltier cell to control the \textit{in-situ} temperature $T$ and thus control the level of activity of the worms.
\textbf{b} Image of a typical \textit{T. Tubifex} worm of length $L$ and width $W$. Scale bar represents 3 mm.
\textbf{c} Shear-thinning curves (shear viscosity $\eta_{s}$ as a function of shear rate $\dot{\gamma}$) of the polymer-like solution ($C$=0.4~g/mL) at different levels of activity, as tuned by adding 5\% of alcohol and varying the temperature ($T$=5, 20, 30$\degree$C). Solid lines are fits to the Cross Model (Eq.~\ref{eq2}).}
\label{F1}
\end{figure*}

\begin{figure*}
\includegraphics[width=16cm]{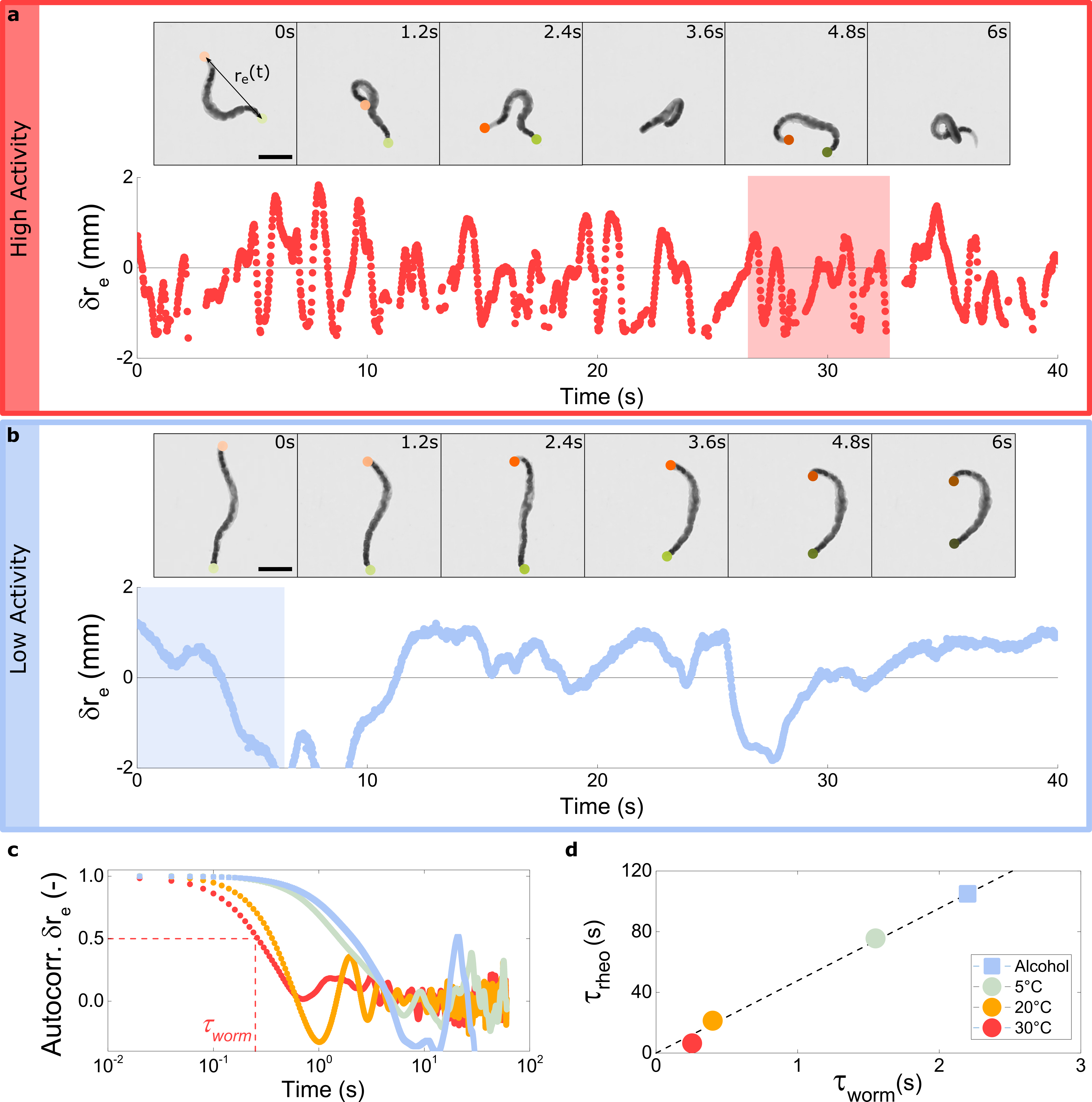}
\caption{\textbf{$\vert$ Microscopic dynamics of a single worm.}
\textbf{a} Sequence of images (initially recorded at a rate of 50 frames per second) of a single worm at a high level of activity in water at $T=30 \degree C$ constrained to move in a quasi-two-dimensional thermo-controlled aquarium (top), and corresponding plot of the fluctuation of the end-to-end distance $\delta {r}_e$ with respect to its averaged value versus time (bottom). Scale bar in the sequence of images represents 2mm. The red shaded area in the plot indicates the 6-second time period during which the sequence of images shown in the top panel were recorded.
\textbf{b} Similar analysis as in panel \textbf{a} for the same worm at a lower level of activity, achieved by lowering the temperature to $T=20 \degree C$ and exposing the worm to 5\% of alcohol.
\textbf{c} Autocorrelation function of $\delta(r_e)$ measured at different levels of activity (\textit{i.e.} different temperatures $T=5,20,30 \degree C$ and in the presence of alcohol). From these graphs, the (microscopic) characteristic time $\tau_{\rm{worm}}$ of a single worm was determined as indicated by the dotted lines.
\textbf{d} Characteristic time of the active solution as deduced from rheology $\tau_{\rm{rheo}}$ plotted against the (microscopic) characteristic time $\tau_{\rm{worm}}$ of a single worm at four levels of activity. The dashed line is a linear fit of the data.} 
\label{F2}
\end{figure*}

We qualitatively study the effect of activity on the shear thinning of the living polymers by performing rheology experiments in a custom-designed plate-plate geometry (Fig.~\ref{F1}, see Methods \& Supplementary Information section I.2). A known mass of Tubifex is mixed with different amounts of (tap) water to adjust the polymer concentration, and put in a beaker with a rough bottom that fits on the rheometer. By inserting an equally rough top plate that just fits into the beaker, we create a plate-plate geometry in which the living polymers are confined and rheology experiments can be performed. The side of the beaker is transparent, allowing us visually confirm that the flows are homogeneous and there is no apparent wall slip (Supplementary Video 1). The setup as a whole is thermostated using Peltier elements, allowing us to control the polymer activity by changing the temperature $T$ (Fig.~\ref{F2}). An efficient way to suppress the activity of the worms is to add 5\% alcohol to the water, which causes almost all of the activity to cease \cite{Gilbertson2016}. This is reversible: if the alcohol is rinsed away using tap water, the activity returns (Supplementary Video 2). In Fig.~\ref{F1}c we compare the rheology of a solution of active worms ($C$=0.2~g/mL, $T$=20$\degree$C) to that of the same solution rendered inactive by adding 5\% alcohol or lowering the temperature. We find that the shear-thinning behaviour is strongly attenuated by activity. The shear thinning power-law exponent changes from  roughly $0$ for the inactive solution to $0.3$ due to activation of the living polymers.  The increased activity also results in lowering of the zero-shear viscosity, effectively flattening the rheology curve.\\

To quantitatively understand these results, we first characterize the activity of the living polymers by investigating the dynamics of single worms. Similarly to what one would do for a normal polymer undergoing thermal fluctuations, we quantify the activity of a typical worm by determining the time scale of its shape fluctuations as a function of the level of activity. This is done by taking an image sequence of a single, isolated Tubifex worm at various temperatures $T$. Fig.~\ref{F2}a,b shows six subsequent images of a typical worm's shape fluctuations in a quasi-two dimensional space (see Methods; Supplementary Information section I.3) for different levels of activity. We quantify the time-dependent variations in the polymer's end-to-end distance $r_e(t)$ as follows:

\begin{equation}
     \delta r_{e}(t) = r_{e}(t)-\langle{r_{e}}\rangle_{t}.
\label{eq1}
\end{equation}

Fig.~\ref{F2}a,b shows typical traces of $\delta r_{e}(t)$ at high ($T=30\degree$C) and low activity (addition of alcohol). Next, we calculate the autocorrelation function:

\begin{equation}
      g(t) = \frac{\langle{\delta r_{e}(t)\delta r_{e}(t+\tau)}\rangle_{t}}{\langle{\delta r_{e}(t)^{2}}\rangle_{t}} .
\label{eq2}
\end{equation}

The (microscopic) characteristic time of the fluctuations $\tau_{\rm{worm}}$ is then determined
from the half-decay time of the autocorrelation function, as shown in Fig.~\ref{F2}c.
We find, as expected, that decreasing the temperature or adding alcohol strongly decreases the activity as quantified by the characteristic time, which increases from 0.26 s at 30$\degree$C to 2.20 s in the presence of alcohol.\\

We can now quantify the shear thinning for different activities. As shown in Fig~\ref{F1}, increasing the shear rate is observed to result in a constant-viscosity Newtonian plateau, followed by power-law type shear thinning. Similarly to usual polymer solutions or melts we therefore analyze these results using a classical polymer rheological model based on the Cross equation \cite{Bird1987}:

\begin{equation}
    \eta_{s}=\frac{\eta_{s,0}}{1+(\dot{\gamma} \tau_{rheo})^{(1-n)}} ,
\label{eq3}
\end{equation}

where $\eta_{s,0}$ is the characteristic zero-shear rate viscosity, $\tau_{\rm{rheo}}$ the average relaxation time (whose inverse corresponds roughly to the onset shear rate for shear thinning) and $n$ the shear thinning exponent that describes the slope of $\eta_{s} / \eta_{s,0}$ in the high-shear-rate power-law region. 
     
As for thermally activated systems, we find for our active system that $\tau_{\rm{rheo}}$ depends on the activity. As shown in Fig.~\ref{F2}d, there is a simple linear relation between the relaxation time obtained from rheology experiments (Fig.~\ref{F1}) and the characteristic time of the microscopic conformational fluctuations of an individual worm (Fig.~\ref{F2}); the slower the microscopic dynamics of the worms, the larger the average relaxation time scale of the ensemble of worms. The observation that the macroscopic rheology times are significantly higher than the microscopic times is in agreement with findings for usual polymers: the onset of shear thinning in the rheology is usually assumed to probe the longest relaxation time in the system \cite{Entov1997,Anna2001}.

For shear thinning of regular polymers, the usual interpretation of shear thinning is that, for low shear rates, the flow is not fast enough to orient the polymer coils that randomize their orientation due to thermal fluctuations. When the shear rate exceeds the inverse of the characteristic time of these thermal fluctuations, the polymers become oriented and a shear-thinning behaviour is observed. Our worms do not exhibit thermal fluctuations but do perform randomizing fluctuations. The rheology suggests that, to first order, the living polymers behave similar to classical polymers, with the worms' activity having a similar orientational randomizing effect as thermal fluctuations.\\

\begin{figure*}
\includegraphics[width=17cm]{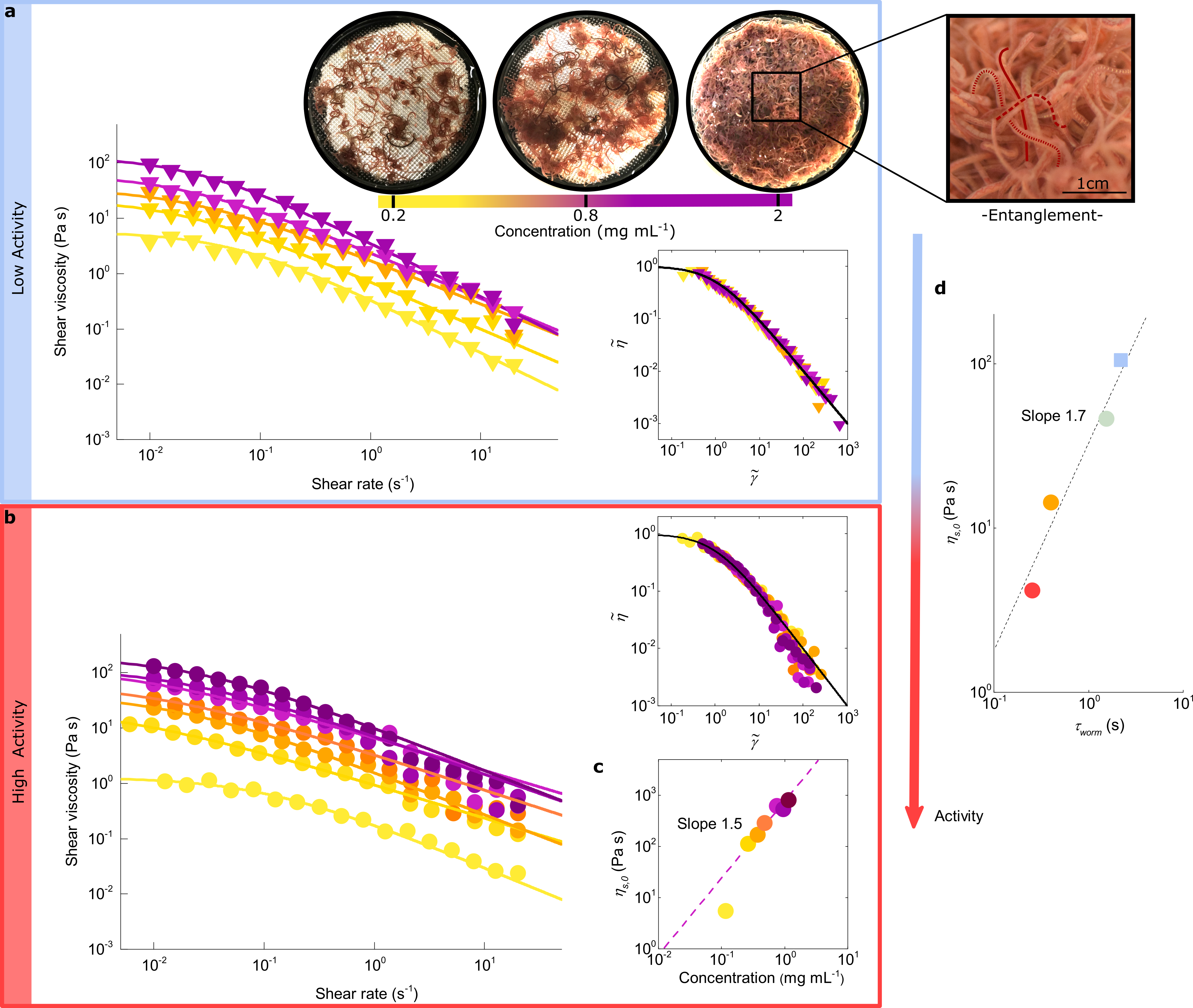}
\caption{\textbf{$\vert$ Effect of the concentration on living polymer rheology.}
\textbf{a,b} Shear viscosity as a function of the concentration of worms in solution at a low ((a): $T=0\degree C$) and high ((b): $T=20\degree C$) level of activity. Insets: As main graph but normalized by the fitting parameters of the Cross model (continuous lines): $\tilde{\eta}=\eta_{s}/ \eta_{s,0}$ and $\tilde{\gamma}=(\dot{\gamma} \tau_{rheo})^{(1-n)}$. The sequence of images shows top views of the container holding typical Tubifex solutions at different concentrations from low (yellow, left) to high (purple, right). A closer view emphasizes the entangled state of the worms at high concentration.  
\textbf{c} Zero-shear viscosity $\eta_{s,0}$ as function of living polymer concentration at a high level of activity ($T=20\degree C$). Dashed line has slope 1.5.
\textbf{d} Effect of activity on the zero-shear viscosity as function of activity as quantified by the (microscopic) characteristic time of the worms ($\tau_{\rm{worm}}$) at a polymer concentration $C$=0.2~g/mL. Dashed line has slope 1.7.}
\label{F3}
\end{figure*}

To see how far the similarities with regular polymer extend, we also investigate the effect of the living polymer concentration on the shear viscosity for two levels of activity. At low activity ($T=0\degree$C, Fig.~\ref{F3}a), the Cross model is sufficient to describe the flow curves for all concentrations (Inset Fig.~\ref{F3}a). These results are at least qualitatively similar to what one would expect for regular (i.e., non-active) polymers: when the concentration increases, the zero-shear viscosity plateau becomes higher, the onset shear rate for shear thinning smaller and the shear-thinning more pronounced. When the solution is more active ($T=20\degree$C, Fig.~\ref{F3}b), similar trends are observed, but at high shear rates deviations from the simple Cross model become apparent (Inset Fig.~\ref{F3}b). It appears that, in this regime, the interaction between living polymers and the flow is more complicated than for regular polymers: the flow is more efficient at orienting living polymers than conventional ones.

We also observe anomalies in the zero-shear viscosity as a function of the living polymer concentration (Fig.~\ref{F3}c). For regular polymers, the scaling with concentration has been much discussed \cite{Heo2005,Vega2004}: simple models give a power-law dependence with an exponent around 3 for the increase of the zero-shear viscosity with concentration. Detailed experiments and more sophisticated theories give an exponent that is slightly higher than 3. Our experiments however reveal a much weaker dependence on concentration, with a power-law exponent around 1.5. In Fig.~\ref{F3}d we show that the zero-shear viscosity is also strongly influenced by polymer activity by comparing it to the characteristic time of an individual worm's fluctuations: the lower the activity, the higher the zero-shear viscosity. The data suggest a power-law relation between the zero-shear viscosity and $\tau_{\rm worm}$, with an exponent of about 1.7.\\ 

Gompper, Winkler and collaborators \cite{Winkler2017,Martin-Gomez2018, Martin-Gomez2019} performed an analytical study of the shear thinning of active polymers for a polymer system consisting of active Brownian particles, meaning that each monomer has a random direction of self-propulsion. Interestingly, their model predicts the opposite of what is observed here, namely that increasing the polymers' activity increases the shear thinning. The discrepancy is likely because their analysis is based on the bead-spring model of connected active particles but may not apply to active worm-like chains. Models that consider the polymers to be tangentially driven \cite{Marchetti2013,Isele-Holder2015} appear to be more suitable to our specific system of living polymers. Indeed, some of the qualitative features described in \cite{Martin-Gomez2019} are very similar to the physical behaviour of our worms. For these models, the rheological implications have not been investigated yet. The similarities as well as differences we report between our system of actively driven polymer-like living worms and well-known polymer solutions that undergo thermal fluctuations invite further scrutiny, opening the new research field of `living polymers'.\\

\textbf{Methods}\\
\\
\textbf{Rheology.} The measurements were carried out using a custom-designed plate-plate geometry adapted from a commercial Anton Paar MCR 302 rheometer. The top and bottom plates (radius $R$=25~mm) surfaces of the geometries have been patterned with pyramidal shapes (see Supplementary Figure 4) of 1~mm large and height (on the order of the diameter of the worms). In that way, the roughness of the top and bottom plates allows to efficiently suppress wall slip effects. The gap of the geometry $h$ is set as the water top level is anchored on the half-height of the upper geometry (see Supplementary Information section I.2). The homogeneity of the flow can be checked during the measurement via the transparent glass window used as lateral container (See Supplementary Video 1) for the active polymer-like solution. The temperature control of the active polymer-like solution is assured by means of a Peltier cell (Anton Paar, P-PTD200/AIR-SN81010431) onto which our custom-designed metal bottom geometry fits, allowing for heat conduction. All shear measurements were carried out two times and on different batches of Tubifex worms to assure reproducibility, and show good agreement in all cases. The total time of the rheology measurement is nevertheless constrained to 2 minutes in water to avoid inhomogeneity of the active system that sets in at longer times: the active systems spontaneously aggregate for biological reasons \cite{Walker1970}.\\

\textbf{Image processing for active polymer-like dynamics.}
The characterization of the activity of the living polymer was investigated by looking at the dynamics of single worms in a quasi-two-dimensional thermo-controlled aquarium designed for this purpose (See Supplementary Information section I.3). The dynamics of the worms are recorded with a Nikon 5300 camera equipped with a macro-lens (Sigma 105~mm 1:2.8) at a frame rate of 50 frames per second. The detection and tracking in time of the end-to-end distance $r_{e}$ of the worms was assured using a home-made algorithm developed in Matlab. The end-to-end distance was extracted from the different processing stages \cite{Brangwynne2007}: thresholding the raw picture, skeletonization and detection of the two end-tips of the skeleton. The home-made algorithm allowed us to accurately record the end-to-end distance of the active polymer as a function of time (See Supplementary Information section I.3). From this information we further deduced the (microscopic) characteristic time of the fluctuation $\tau_{\rm{worm}}$ (see main text).

\textbf{Acknowledgements.} We thank the workshop of the University of Amsterdam for their skilful technical assistance. We are grateful to the Aquarium Holgen for providing fresh batches of T. Tubifex worms.\\ 
A.D acknowledges the funding from the European Union’s Horizon 2020 research and innovation program under the Individual Marie Skłodowska-Curie fellowship grant agreement number 798455.\\

\textbf{Author Contributions.} S. Woutersen and D. Bonn conceived the presented idea. All authors performed the experiments, discussed the results and contributed to the final manuscript.\\

\textbf{Competing interests} A. Deblais, S. Woutersen and D. Bonn declare no competing interests.\\

\textbf{Correspondence and requests for materials} should be addressed to S. Woutersen, A. Deblais or D. Bonn.\\
-A.Deblais@uva.nl\\
-S.Woutersen@uva.nl\\
-D.Bonn@uva.nl\\

\begin{thebibliography}{32}%
\makeatletter
\providecommand \@ifxundefined [1]{%
 \@ifx{#1\undefined}
}%
\providecommand \@ifnum [1]{%
 \ifnum #1\expandafter \@firstoftwo
 \else \expandafter \@secondoftwo
 \fi
}%
\providecommand \@ifx [1]{%
 \ifx #1\expandafter \@firstoftwo
 \else \expandafter \@secondoftwo
 \fi
}%
\providecommand \natexlab [1]{#1}%
\providecommand \enquote  [1]{``#1''}%
\providecommand \bibnamefont  [1]{#1}%
\providecommand \bibfnamefont [1]{#1}%
\providecommand \citenamefont [1]{#1}%
\providecommand \href@noop [0]{\@secondoftwo}%
\providecommand \href [0]{\begingroup \@sanitize@url \@href}%
\providecommand \@href[1]{\@@startlink{#1}\@@href}%
\providecommand \@@href[1]{\endgroup#1\@@endlink}%
\providecommand \@sanitize@url [0]{\catcode `\\12\catcode `\$12\catcode
  `\&12\catcode `\#12\catcode `\^12\catcode `\_12\catcode `\%12\relax}%
\providecommand \@@startlink[1]{}%
\providecommand \@@endlink[0]{}%
\providecommand \url  [0]{\begingroup\@sanitize@url \@url }%
\providecommand \@url [1]{\endgroup\@href {#1}{\urlprefix }}%
\providecommand \urlprefix  [0]{URL }%
\providecommand \Eprint [0]{\href }%
\providecommand \doibase [0]{http://dx.doi.org/}%
\providecommand \selectlanguage [0]{\@gobble}%
\providecommand \bibinfo  [0]{\@secondoftwo}%
\providecommand \bibfield  [0]{\@secondoftwo}%
\providecommand \translation [1]{[#1]}%
\providecommand \BibitemOpen [0]{}%
\providecommand \bibitemStop [0]{}%
\providecommand \bibitemNoStop [0]{.\EOS\space}%
\providecommand \EOS [0]{\spacefactor3000\relax}%
\providecommand \BibitemShut  [1]{\csname bibitem#1\endcsname}%
\let\auto@bib@innerbib\@empty
%</preamble>
\bibitem [{\citenamefont {{De Gennes}}(1979)}]{DeGennes1979}%
  \BibitemOpen
  \bibfield  {author} {\bibinfo {author} {\bibfnamefont {P.~G.}\ \bibnamefont
  {{De Gennes}}},\ }\href@noop {} {\emph {\bibinfo {title} {{Scaling concepts
  in polymer physics}}}},\ edited by\ \bibinfo {editor} {\bibfnamefont
  {C.}~\bibnamefont {university press}}\ (\bibinfo  {publisher} {London, UK:
  Cornell University Press},\ \bibinfo {year} {1979})\BibitemShut {NoStop}%
\bibitem [{\citenamefont {Doi}(1986)}]{Doi1986}%
  \BibitemOpen
  \bibfield  {author} {\bibinfo {author} {\bibfnamefont {M.}~\bibnamefont
  {Doi}},\ }\href@noop {} {\bibfield  {journal} {\bibinfo  {journal} {Clarendon
  Press, Oxford}\ } (\bibinfo {year} {1986})}\BibitemShut {NoStop}%
\bibitem [{\citenamefont {Munk}\ and\ \citenamefont
  {Aminabhavi}(1989)}]{Munk1989}%
  \BibitemOpen
  \bibfield  {author} {\bibinfo {author} {\bibfnamefont {P.}~\bibnamefont
  {Munk}}\ and\ \bibinfo {author} {\bibfnamefont {T.~M.}\ \bibnamefont
  {Aminabhavi}},\ }\href@noop {} {\emph {\bibinfo {title} {{Introduction to
  macromolecular science}}}}\ (\bibinfo  {publisher} {Wiley New York},\
  \bibinfo {year} {1989})\BibitemShut {NoStop}%
\bibitem [{\citenamefont {LeDuc}\ \emph {et~al.}(1999)\citenamefont {LeDuc},
  \citenamefont {Haber}, \citenamefont {Bao},\ and\ \citenamefont
  {Wirtz}}]{LeDuc1999}%
  \BibitemOpen
  \bibfield  {author} {\bibinfo {author} {\bibfnamefont {P.}~\bibnamefont
  {LeDuc}}, \bibinfo {author} {\bibfnamefont {C.}~\bibnamefont {Haber}},
  \bibinfo {author} {\bibfnamefont {G.}~\bibnamefont {Bao}}, \ and\ \bibinfo
  {author} {\bibfnamefont {D.}~\bibnamefont {Wirtz}},\ }\href {\doibase
  10.1038/21148} {\bibfield  {journal} {\bibinfo  {journal} {Nature}\ }\textbf
  {\bibinfo {volume} {399}},\ \bibinfo {pages} {564} (\bibinfo {year}
  {1999})}\BibitemShut {NoStop}%
\bibitem [{\citenamefont {Perazzo}\ \emph {et~al.}(2017)\citenamefont
  {Perazzo}, \citenamefont {Nunes}, \citenamefont {Guido},\ and\ \citenamefont
  {Stone}}]{Perazzo2017}%
  \BibitemOpen
  \bibfield  {author} {\bibinfo {author} {\bibfnamefont {A.}~\bibnamefont
  {Perazzo}}, \bibinfo {author} {\bibfnamefont {J.~K.}\ \bibnamefont {Nunes}},
  \bibinfo {author} {\bibfnamefont {S.}~\bibnamefont {Guido}}, \ and\ \bibinfo
  {author} {\bibfnamefont {H.~A.}\ \bibnamefont {Stone}},\ }\href {\doibase
  10.1073/pnas.1710927114} {\bibfield  {journal} {\bibinfo  {journal}
  {Proceedings of the National Academy of Sciences of the United States of
  America}\ }\textbf {\bibinfo {volume} {114}},\ \bibinfo {pages} {E8557}
  (\bibinfo {year} {2017})}\BibitemShut {NoStop}%
\bibitem [{\citenamefont {Liverpool}\ \emph {et~al.}(2001)\citenamefont
  {Liverpool}, \citenamefont {Maggs},\ and\ \citenamefont
  {Ajdari}}]{Liverpool2001}%
  \BibitemOpen
  \bibfield  {author} {\bibinfo {author} {\bibfnamefont {T.~B.}\ \bibnamefont
  {Liverpool}}, \bibinfo {author} {\bibfnamefont {A.~C.}\ \bibnamefont
  {Maggs}}, \ and\ \bibinfo {author} {\bibfnamefont {A.}~\bibnamefont
  {Ajdari}},\ }\href {\doibase 10.1103/PhysRevLett.86.4171} {\bibfield
  {journal} {\bibinfo  {journal} {Physical Review Letters}\ }\textbf {\bibinfo
  {volume} {86}},\ \bibinfo {pages} {4171} (\bibinfo {year}
  {2001})}\BibitemShut {NoStop}%
\bibitem [{\citenamefont {Marchetti}\ \emph {et~al.}(2013)\citenamefont
  {Marchetti}, \citenamefont {Joanny}, \citenamefont {Ramaswamy}, \citenamefont
  {Liverpool}, \citenamefont {Prost}, \citenamefont {Rao},\ and\ \citenamefont
  {Simha}}]{Marchetti2013}%
  \BibitemOpen
  \bibfield  {author} {\bibinfo {author} {\bibfnamefont {M.~C.}\ \bibnamefont
  {Marchetti}}, \bibinfo {author} {\bibfnamefont {J.~F.}\ \bibnamefont
  {Joanny}}, \bibinfo {author} {\bibfnamefont {S.}~\bibnamefont {Ramaswamy}},
  \bibinfo {author} {\bibfnamefont {T.~B.}\ \bibnamefont {Liverpool}}, \bibinfo
  {author} {\bibfnamefont {J.}~\bibnamefont {Prost}}, \bibinfo {author}
  {\bibfnamefont {M.}~\bibnamefont {Rao}}, \ and\ \bibinfo {author}
  {\bibfnamefont {R.~A.}\ \bibnamefont {Simha}},\ }\href {\doibase
  10.1103/RevModPhys.85.1143} {\bibfield  {journal} {\bibinfo  {journal}
  {Reviews of Modern Physics}\ }\textbf {\bibinfo {volume} {85}},\ \bibinfo
  {pages} {1143} (\bibinfo {year} {2013})}\BibitemShut {NoStop}%
\bibitem [{\citenamefont {Winkler}\ \emph {et~al.}(2017)\citenamefont
  {Winkler}, \citenamefont {Elgeti},\ and\ \citenamefont
  {Gompper}}]{Winkler2017}%
  \BibitemOpen
  \bibfield  {author} {\bibinfo {author} {\bibfnamefont {R.~G.}\ \bibnamefont
  {Winkler}}, \bibinfo {author} {\bibfnamefont {J.}~\bibnamefont {Elgeti}}, \
  and\ \bibinfo {author} {\bibfnamefont {G.}~\bibnamefont {Gompper}},\ }\href
  {\doibase 10.7566/JPSJ.86.101014} {\bibfield  {journal} {\bibinfo  {journal}
  {Journal of the Physical Society of Japan}\ }\textbf {\bibinfo {volume}
  {86}},\ \bibinfo {pages} {1} (\bibinfo {year} {2017})}\BibitemShut {NoStop}%
\bibitem [{\citenamefont {Mart{\'{i}}n-G{\'{o}}mez}\ \emph
  {et~al.}(2018)\citenamefont {Mart{\'{i}}n-G{\'{o}}mez}, \citenamefont
  {Gompper},\ and\ \citenamefont {Winkler}}]{Martin-Gomez2018}%
  \BibitemOpen
  \bibfield  {author} {\bibinfo {author} {\bibfnamefont {A.}~\bibnamefont
  {Mart{\'{i}}n-G{\'{o}}mez}}, \bibinfo {author} {\bibfnamefont
  {G.}~\bibnamefont {Gompper}}, \ and\ \bibinfo {author} {\bibfnamefont
  {R.~G.}\ \bibnamefont {Winkler}},\ }\href {\doibase 10.3390/polym10080837}
  {\bibfield  {journal} {\bibinfo  {journal} {Polymers}\ }\textbf {\bibinfo
  {volume} {10}},\ \bibinfo {pages} {60} (\bibinfo {year} {2018})}\BibitemShut
  {NoStop}%
\bibitem [{\citenamefont {Mart{\'{i}}n-G{\'{o}}mez}\ \emph
  {et~al.}(2019)\citenamefont {Mart{\'{i}}n-G{\'{o}}mez}, \citenamefont
  {Eisenstecken}, \citenamefont {Gompper},\ and\ \citenamefont
  {Winkler}}]{Martin-Gomez2019}%
  \BibitemOpen
  \bibfield  {author} {\bibinfo {author} {\bibfnamefont {A.}~\bibnamefont
  {Mart{\'{i}}n-G{\'{o}}mez}}, \bibinfo {author} {\bibfnamefont
  {T.}~\bibnamefont {Eisenstecken}}, \bibinfo {author} {\bibfnamefont
  {G.}~\bibnamefont {Gompper}}, \ and\ \bibinfo {author} {\bibfnamefont
  {R.~G.}\ \bibnamefont {Winkler}},\ }\href {\doibase 10.1039/c9sm00391f}
  {\bibfield  {journal} {\bibinfo  {journal} {Soft Matter}\ }\textbf {\bibinfo
  {volume} {15}},\ \bibinfo {pages} {3957} (\bibinfo {year}
  {2019})}\BibitemShut {NoStop}%
\bibitem [{\citenamefont {Lazim}\ and\ \citenamefont
  {Learner}(1986)}]{Lazim1986}%
  \BibitemOpen
  \bibfield  {author} {\bibinfo {author} {\bibfnamefont {M.~N.}\ \bibnamefont
  {Lazim}}\ and\ \bibinfo {author} {\bibfnamefont {M.~A.}\ \bibnamefont
  {Learner}},\ }\href {\doibase 10.1111/j.1600-0587.1986.tb01208.x} {\bibfield
  {journal} {\bibinfo  {journal} {Ecography}\ }\textbf {\bibinfo {volume}
  {9}},\ \bibinfo {pages} {185} (\bibinfo {year} {1986})}\BibitemShut {NoStop}%
\bibitem [{\citenamefont {Zhang}\ \emph {et~al.}(2010)\citenamefont {Zhang},
  \citenamefont {Be'er}, \citenamefont {Florin},\ and\ \citenamefont
  {Swinney}}]{Zhang2010}%
  \BibitemOpen
  \bibfield  {author} {\bibinfo {author} {\bibfnamefont {H.-P.}\ \bibnamefont
  {Zhang}}, \bibinfo {author} {\bibfnamefont {A.}~\bibnamefont {Be'er}},
  \bibinfo {author} {\bibfnamefont {E.-L.}\ \bibnamefont {Florin}}, \ and\
  \bibinfo {author} {\bibfnamefont {H.~L.}\ \bibnamefont {Swinney}},\
  }\href@noop {} {\bibfield  {journal} {\bibinfo  {journal} {Proceedings of the
  National Academy of Sciences}\ }\textbf {\bibinfo {volume} {107}},\ \bibinfo
  {pages} {13626} (\bibinfo {year} {2010})}\BibitemShut {NoStop}%
\bibitem [{\citenamefont {Fodor}\ and\ \citenamefont {{Cristina
  Marchetti}}(2018)}]{Fodor2018}%
  \BibitemOpen
  \bibfield  {author} {\bibinfo {author} {\bibfnamefont {{\'{E}}.}~\bibnamefont
  {Fodor}}\ and\ \bibinfo {author} {\bibfnamefont {M.}~\bibnamefont {{Cristina
  Marchetti}}},\ }\href {\doibase 10.1016/j.physa.2017.12.137} {\bibfield
  {journal} {\bibinfo  {journal} {Physica A: Statistical Mechanics and its
  Applications}\ }\textbf {\bibinfo {volume} {504}},\ \bibinfo {pages} {106}
  (\bibinfo {year} {2018})}\BibitemShut {NoStop}%
\bibitem [{\citenamefont {Chen}\ \emph {et~al.}(2007)\citenamefont {Chen},
  \citenamefont {Lau}, \citenamefont {Hough}, \citenamefont {Islam},
  \citenamefont {Goulian}, \citenamefont {Lubensky},\ and\ \citenamefont
  {Yodh}}]{Chen2007}%
  \BibitemOpen
  \bibfield  {author} {\bibinfo {author} {\bibfnamefont {D.~T.~N.}\
  \bibnamefont {Chen}}, \bibinfo {author} {\bibfnamefont {A.~W.~C.}\
  \bibnamefont {Lau}}, \bibinfo {author} {\bibfnamefont {L.~A.}\ \bibnamefont
  {Hough}}, \bibinfo {author} {\bibfnamefont {M.~F.}\ \bibnamefont {Islam}},
  \bibinfo {author} {\bibfnamefont {M.}~\bibnamefont {Goulian}}, \bibinfo
  {author} {\bibfnamefont {T.~C.}\ \bibnamefont {Lubensky}}, \ and\ \bibinfo
  {author} {\bibfnamefont {A.~G.}\ \bibnamefont {Yodh}},\ }\href {\doibase
  10.1103/PhysRevLett.99.148302} {\bibfield  {journal} {\bibinfo  {journal}
  {Physical Review Letters}\ }\textbf {\bibinfo {volume} {148302}},\ \bibinfo
  {pages} {1} (\bibinfo {year} {2007})}\BibitemShut {NoStop}%
\bibitem [{\citenamefont {Sokolov}\ and\ \citenamefont
  {Aranson}(2009)}]{Sokolov2009}%
  \BibitemOpen
  \bibfield  {author} {\bibinfo {author} {\bibfnamefont {A.}~\bibnamefont
  {Sokolov}}\ and\ \bibinfo {author} {\bibfnamefont {I.~S.}\ \bibnamefont
  {Aranson}},\ }\href {\doibase 10.1103/PhysRevLett.103.148101} {\bibfield
  {journal} {\bibinfo  {journal} {Physical Review Letters}\ }\textbf {\bibinfo
  {volume} {103}},\ \bibinfo {pages} {2} (\bibinfo {year} {2009})}\BibitemShut
  {NoStop}%
\bibitem [{\citenamefont {Giomi}\ \emph {et~al.}(2010)\citenamefont {Giomi},
  \citenamefont {Liverpool},\ and\ \citenamefont {Marchetti}}]{Giomi2010}%
  \BibitemOpen
  \bibfield  {author} {\bibinfo {author} {\bibfnamefont {L.}~\bibnamefont
  {Giomi}}, \bibinfo {author} {\bibfnamefont {T.~B.}\ \bibnamefont
  {Liverpool}}, \ and\ \bibinfo {author} {\bibfnamefont {M.~C.}\ \bibnamefont
  {Marchetti}},\ }\href {\doibase 10.1103/PhysRevE.81.051908} {\bibfield
  {journal} {\bibinfo  {journal} {Physical Review Letters}\ ,\ \bibinfo {pages}
  {1}} (\bibinfo {year} {2010})}\BibitemShut {NoStop}%
\bibitem [{\citenamefont {Koch}\ and\ \citenamefont
  {Subramanian}(2011)}]{Koch2011}%
  \BibitemOpen
  \bibfield  {author} {\bibinfo {author} {\bibfnamefont {D.~L.}\ \bibnamefont
  {Koch}}\ and\ \bibinfo {author} {\bibfnamefont {G.}~\bibnamefont
  {Subramanian}},\ }\href {\doibase 10.1146/annurev-fluid-121108-145434}
  {\bibfield  {journal} {\bibinfo  {journal} {Annual Review of Fluid
  Mechanics}\ }\textbf {\bibinfo {volume} {43}},\ \bibinfo {pages} {637}
  (\bibinfo {year} {2011})}\BibitemShut {NoStop}%
\bibitem [{\citenamefont {Zhou}\ \emph {et~al.}(2013)\citenamefont {Zhou},
  \citenamefont {Martinez},\ and\ \citenamefont {Fredberg}}]{Zhou2013}%
  \BibitemOpen
  \bibfield  {author} {\bibinfo {author} {\bibfnamefont {E.~H.}\ \bibnamefont
  {Zhou}}, \bibinfo {author} {\bibfnamefont {F.~D.}\ \bibnamefont {Martinez}},
  \ and\ \bibinfo {author} {\bibfnamefont {J.~J.}\ \bibnamefont {Fredberg}},\
  }\href {\doibase 10.1038/nmat3574} {\bibfield  {journal} {\bibinfo  {journal}
  {Nature Materials}\ }\textbf {\bibinfo {volume} {12}},\ \bibinfo {pages}
  {184} (\bibinfo {year} {2013})}\BibitemShut {NoStop}%
\bibitem [{\citenamefont {L{\'{o}}pez}\ \emph {et~al.}(2015)\citenamefont
  {L{\'{o}}pez}, \citenamefont {Gachelin}, \citenamefont {Douarche},
  \citenamefont {Auradou},\ and\ \citenamefont {Cl{\'{e}}ment}}]{Lopez2015}%
  \BibitemOpen
  \bibfield  {author} {\bibinfo {author} {\bibfnamefont {H.~M.}\ \bibnamefont
  {L{\'{o}}pez}}, \bibinfo {author} {\bibfnamefont {J.}~\bibnamefont
  {Gachelin}}, \bibinfo {author} {\bibfnamefont {C.}~\bibnamefont {Douarche}},
  \bibinfo {author} {\bibfnamefont {H.}~\bibnamefont {Auradou}}, \ and\
  \bibinfo {author} {\bibfnamefont {E.}~\bibnamefont {Cl{\'{e}}ment}},\ }\href
  {\doibase 10.1103/PhysRevLett.115.028301} {\bibfield  {journal} {\bibinfo
  {journal} {Physical Review Letters}\ }\textbf {\bibinfo {volume} {115}},\
  \bibinfo {pages} {1} (\bibinfo {year} {2015})}\BibitemShut {NoStop}%
\bibitem [{\citenamefont {Tennenbaum}\ \emph {et~al.}(2016)\citenamefont
  {Tennenbaum}, \citenamefont {Liu}, \citenamefont {Hu},\ and\ \citenamefont
  {Fernandez-nieves}}]{Tennenbaum2016}%
  \BibitemOpen
  \bibfield  {author} {\bibinfo {author} {\bibfnamefont {M.}~\bibnamefont
  {Tennenbaum}}, \bibinfo {author} {\bibfnamefont {Z.}~\bibnamefont {Liu}},
  \bibinfo {author} {\bibfnamefont {D.}~\bibnamefont {Hu}}, \ and\ \bibinfo
  {author} {\bibfnamefont {A.}~\bibnamefont {Fernandez-nieves}},\ }\href@noop
  {} {\bibfield  {journal} {\bibinfo  {journal} {Nature Materials}\ }\textbf
  {\bibinfo {volume} {15}} (\bibinfo {year} {2016})}\BibitemShut {NoStop}%
\bibitem [{\citenamefont {Hatwalne}\ \emph {et~al.}(2004)\citenamefont
  {Hatwalne}, \citenamefont {Ramaswamy}, \citenamefont {Rao},\ and\
  \citenamefont {Simha}}]{Hatwalne2004}%
  \BibitemOpen
  \bibfield  {author} {\bibinfo {author} {\bibfnamefont {Y.}~\bibnamefont
  {Hatwalne}}, \bibinfo {author} {\bibfnamefont {S.}~\bibnamefont {Ramaswamy}},
  \bibinfo {author} {\bibfnamefont {M.}~\bibnamefont {Rao}}, \ and\ \bibinfo
  {author} {\bibfnamefont {R.~A.}\ \bibnamefont {Simha}},\ }\href {\doibase
  10.1103/PhysRevLett.92.118101} {\bibfield  {journal} {\bibinfo  {journal}
  {Physical Review Letters}\ }\textbf {\bibinfo {volume} {92}},\ \bibinfo
  {pages} {1} (\bibinfo {year} {2004})}\BibitemShut {NoStop}%
\bibitem [{\citenamefont {Linden}\ \emph {et~al.}(2019)\citenamefont {Linden},
  \citenamefont {Alexander}, \citenamefont {Aarts},\ and\ \citenamefont
  {Dauchot}}]{Linden2019}%
  \BibitemOpen
  \bibfield  {author} {\bibinfo {author} {\bibfnamefont {M.~N. V.~D.}\
  \bibnamefont {Linden}}, \bibinfo {author} {\bibfnamefont {L.~C.}\
  \bibnamefont {Alexander}}, \bibinfo {author} {\bibfnamefont {D.~G. A.~L.}\
  \bibnamefont {Aarts}}, \ and\ \bibinfo {author} {\bibfnamefont
  {O.}~\bibnamefont {Dauchot}},\ }\href@noop {} {\ ,\ \bibinfo {pages} {1}
  (\bibinfo {year} {2019})},\ \Eprint {http://arxiv.org/abs/arXiv:1902.08094v1}
  {arXiv:arXiv:1902.08094v1} \BibitemShut {NoStop}%
\bibitem [{\citenamefont {Geyer}\ \emph {et~al.}(2019)\citenamefont {Geyer},
  \citenamefont {David}, \citenamefont {Tailleur},\ and\ \citenamefont
  {Bartolo}}]{Delphine2019}%
  \BibitemOpen
  \bibfield  {author} {\bibinfo {author} {\bibfnamefont {D.}~\bibnamefont
  {Geyer}}, \bibinfo {author} {\bibfnamefont {M.}~\bibnamefont {David}},
  \bibinfo {author} {\bibfnamefont {J.}~\bibnamefont {Tailleur}}, \ and\
  \bibinfo {author} {\bibfnamefont {D.}~\bibnamefont {Bartolo}},\ }\href
  {\doibase 10.1103/PhysRevX.9.031043} {\bibfield  {journal} {\bibinfo
  {journal} {Physical Review X}\ }\textbf {\bibinfo {volume} {9}},\ \bibinfo
  {pages} {31043} (\bibinfo {year} {2019})},\ \Eprint
  {http://arxiv.org/abs/1903.01134} {arXiv:1903.01134} \BibitemShut {NoStop}%
\bibitem [{\citenamefont {Gilbertson}\ and\ \citenamefont
  {Wyatt}(2016)}]{Gilbertson2016}%
  \BibitemOpen
  \bibfield  {author} {\bibinfo {author} {\bibfnamefont {C.~R.}\ \bibnamefont
  {Gilbertson}}\ and\ \bibinfo {author} {\bibfnamefont {J.~D.}\ \bibnamefont
  {Wyatt}},\ }\href@noop {} {\ \textbf {\bibinfo {volume} {55}},\ \bibinfo
  {pages} {577} (\bibinfo {year} {2016})}\BibitemShut {NoStop}%
\bibitem [{\citenamefont {Bird}\ \emph {et~al.}(1987)\citenamefont {Bird},
  \citenamefont {Armstrong},\ and\ \citenamefont {Hassager}}]{Bird1987}%
  \BibitemOpen
  \bibfield  {author} {\bibinfo {author} {\bibfnamefont {R.~B.}\ \bibnamefont
  {Bird}}, \bibinfo {author} {\bibfnamefont {R.~C.}\ \bibnamefont {Armstrong}},
  \ and\ \bibinfo {author} {\bibfnamefont {O.}~\bibnamefont {Hassager}},\
  }\href@noop {} {\bibfield  {journal} {\bibinfo  {journal} {Dynamics of
  Polymer Liquids}\ }\textbf {\bibinfo {volume} {2}} (\bibinfo {year}
  {1987})}\BibitemShut {NoStop}%
\bibitem [{\citenamefont {Entov}\ and\ \citenamefont
  {Hinch}(1997)}]{Entov1997}%
  \BibitemOpen
  \bibfield  {author} {\bibinfo {author} {\bibfnamefont {V.~M.}\ \bibnamefont
  {Entov}}\ and\ \bibinfo {author} {\bibfnamefont {E.~J.}\ \bibnamefont
  {Hinch}},\ }\href@noop {} {\bibfield  {journal} {\bibinfo  {journal} {Journal
  of Non-Newtonian Fluid Mechanics}\ }\textbf {\bibinfo {volume} {72}},\
  \bibinfo {pages} {31} (\bibinfo {year} {1997})}\BibitemShut {NoStop}%
\bibitem [{\citenamefont {Anna}\ and\ \citenamefont
  {McKinley}(2001)}]{Anna2001}%
  \BibitemOpen
  \bibfield  {author} {\bibinfo {author} {\bibfnamefont {S.~L.}\ \bibnamefont
  {Anna}}\ and\ \bibinfo {author} {\bibfnamefont {G.~H.}\ \bibnamefont
  {McKinley}},\ }\href {\doibase 10.1122/1.1332389} {\bibfield  {journal}
  {\bibinfo  {journal} {Journal of Rheology}\ }\textbf {\bibinfo {volume}
  {45}},\ \bibinfo {pages} {115} (\bibinfo {year} {2001})}\BibitemShut
  {NoStop}%
\bibitem [{\citenamefont {Heo}\ and\ \citenamefont {Larson}(2005)}]{Heo2005}%
  \BibitemOpen
  \bibfield  {author} {\bibinfo {author} {\bibfnamefont {Y.}~\bibnamefont
  {Heo}}\ and\ \bibinfo {author} {\bibfnamefont {R.~G.}\ \bibnamefont
  {Larson}},\ }\href {\doibase 10.1122/1.1993595} {\bibfield  {journal}
  {\bibinfo  {journal} {Journal of Rheology}\ }\textbf {\bibinfo {volume}
  {49}},\ \bibinfo {pages} {1117} (\bibinfo {year} {2005})}\BibitemShut
  {NoStop}%
\bibitem [{\citenamefont {Vega}\ \emph {et~al.}(2004)\citenamefont {Vega},
  \citenamefont {Rastogi}, \citenamefont {Peters},\ and\ \citenamefont
  {Meijer}}]{Vega2004}%
  \BibitemOpen
  \bibfield  {author} {\bibinfo {author} {\bibfnamefont {J.~F.}\ \bibnamefont
  {Vega}}, \bibinfo {author} {\bibfnamefont {S.}~\bibnamefont {Rastogi}},
  \bibinfo {author} {\bibfnamefont {G.~W.~M.}\ \bibnamefont {Peters}}, \ and\
  \bibinfo {author} {\bibfnamefont {H.~E.~H.}\ \bibnamefont {Meijer}},\ }\href
  {\doibase 10.1122/1.1718367} {\bibfield  {journal} {\bibinfo  {journal}
  {Journal of Rheology}\ }\textbf {\bibinfo {volume} {48}},\ \bibinfo {pages}
  {663} (\bibinfo {year} {2004})}\BibitemShut {NoStop}%
\bibitem [{\citenamefont {Isele-Holder}\ \emph {et~al.}(2015)\citenamefont
  {Isele-Holder}, \citenamefont {Elgeti},\ and\ \citenamefont
  {Gompper}}]{Isele-Holder2015}%
  \BibitemOpen
  \bibfield  {author} {\bibinfo {author} {\bibfnamefont {R.~E.}\ \bibnamefont
  {Isele-Holder}}, \bibinfo {author} {\bibfnamefont {J.}~\bibnamefont
  {Elgeti}}, \ and\ \bibinfo {author} {\bibfnamefont {G.}~\bibnamefont
  {Gompper}},\ }\href {\doibase 10.1039/c5sm01683e} {\bibfield  {journal}
  {\bibinfo  {journal} {Soft Matter}\ }\textbf {\bibinfo {volume} {11}},\
  \bibinfo {pages} {7181} (\bibinfo {year} {2015})}\BibitemShut {NoStop}%
\bibitem [{\citenamefont {Walker}(1970)}]{Walker1970}%
  \BibitemOpen
  \bibfield  {author} {\bibinfo {author} {\bibfnamefont {J.~G.}\ \bibnamefont
  {Walker}},\ }\href@noop {} {\bibfield  {journal} {\bibinfo  {journal} {Biol.
  Bull.}\ ,\ \bibinfo {pages} {235}} (\bibinfo {year} {1970})}\BibitemShut
  {NoStop}%
\bibitem [{\citenamefont {Brangwynne}\ \emph {et~al.}(2007)\citenamefont
  {Brangwynne}, \citenamefont {Koenderink}, \citenamefont {Barry},
  \citenamefont {Dogic}, \citenamefont {MacKintosh},\ and\ \citenamefont
  {Weitz}}]{Brangwynne2007}%
  \BibitemOpen
  \bibfield  {author} {\bibinfo {author} {\bibfnamefont {C.~P.}\ \bibnamefont
  {Brangwynne}}, \bibinfo {author} {\bibfnamefont {G.~H.}\ \bibnamefont
  {Koenderink}}, \bibinfo {author} {\bibfnamefont {E.}~\bibnamefont {Barry}},
  \bibinfo {author} {\bibfnamefont {Z.}~\bibnamefont {Dogic}}, \bibinfo
  {author} {\bibfnamefont {F.~C.}\ \bibnamefont {MacKintosh}}, \ and\ \bibinfo
  {author} {\bibfnamefont {D.~A.}\ \bibnamefont {Weitz}},\ }\href {\doibase
  10.1529/biophysj.106.096966} {\bibfield  {journal} {\bibinfo  {journal}
  {Biophysical Journal}\ }\textbf {\bibinfo {volume} {93}},\ \bibinfo {pages}
  {346} (\bibinfo {year} {2007})}\BibitemShut {NoStop}%
\end{thebibliography}
\end{document}